%% file: main.tex
\documentclass[conference]{IEEEtran}

\usepackage[utf8]{inputenc}
\usepackage{todonotes}
\usepackage{paralist}
\usepackage{booktabs}
\usepackage{rotating}
\usepackage[bookmarks=false, hidelinks]{hyperref}
\usepackage{chngpage}
\usepackage{multirow}
\usepackage{amssymb}
\usepackage{amsmath}
\usepackage{caption}
\usepackage{subcaption}
\usepackage{framed}
\usepackage{lipsum}
\usepackage{cite}
\usepackage[numbers, compress]{natbib}
\let\OLDthebibliography\thebibliography
\renewcommand\thebibliography[1]{
  \OLDthebibliography{#1}
  \setlength{\parskip}{0pt}
  \setlength{\itemsep}{0pt plus 0.3ex}
}

\ifCLASSINFOpdf
\else
\fi

\usepackage{balance}

\begin{document}

\title{On the Impact of Semantic Transparency on Understanding and Reviewing Social Goal Models}




\author{\IEEEauthorblockN{Mafalda Santos, Catarina Gralha, Miguel Goulão, João Araújo, Ana Moreira
}
\IEEEauthorblockA{NOVA LINCS, Department of Computer Science
\\
Faculty of Science and Technology, Universidade NOVA de Lisboa
\\
\{mcd.santos, acg.almeida\}@campus.fct.unl.pt, \{mgoul, joao.araujo, amm\}@fct.unl.pt
} 
}

\maketitle

\input{0_abstract}

\IEEEpeerreviewmaketitle

\input{1_introduction}
\input{3_new_iStar_Notation}
\input{4_experiment}
\input{5_execution}
\input{6_analysis}
\input{7_discussion}
\input{8_related_work}

\input{9_conclusions}

\section*{Acknowledgments}
We thank NOVA LINCS UID/CEC/04516/2013 and FCT-MCTES SFRH/BD/108492/2015 for financial support.

\bibliographystyle{IEEEtran}
\bibliography{references}
\balance

\end{document}

%% file: 0_abstract.tex
\begin{abstract}

\emph{Context:} 
\textit{i*} is one of the most influential languages in the Requirements Engineering research community. Perhaps due to its complexity and low adoption in industry, it became a natural candidate for studies aiming at improving its concrete syntax and the stakeholders' ability to correctly interpret \textit{i*} models.
\emph{Objectives:} 
We evaluate 
the impact of semantic transparency on understanding and reviewing \textit{i*} models,
in the presence of a language key.
\emph{Methods:} 
We performed a quasi-experiment comparing  
the standard \textit{i*} concrete syntax with an alternative that has an increased semantic transparency. 
We asked 57 novice participants to perform understanding and reviewing tasks on \textit{i*} models, and measured their \textit{accuracy}, \textit{speed} and \textit{ease}, using metrics of task success, time and effort, collected with eye-tracking and participants' feedback.
\emph{Results:} 
We found no evidence of improved accuracy or speed attributable to the 
alternative concrete syntax. Although participants' perceived ease was similar, they devoted significantly less visual effort to the model and the provided language key, when using the 
alternative concrete syntax. 
\emph{Conclusions:}
The context provided by the model and language key may mitigate the \textit{i*} symbol recognition deficit reported in previous works. However, the 
alternative concrete syntax required a significantly lower visual effort.
\end{abstract}

\begin{IEEEkeywords}
     social goal models, i*, physics of notations, eye-tracking
\end{IEEEkeywords}


%% file: 1_introduction.tex
\section{Introduction}
\label{sec:introduction}

Requirements Engineering (RE) success depends on, among several other factors, the quality of the communication between requirements engineers and other stakeholders. Indeed, communication flaws are among the most frequently reported RE problems that may lead to project failure \cite{Fernandez2016EMSE}. One of the key elements of an effective communication is the language used. Visual notations are often adopted, as they are perceived as more effective for conveying information to non-technical stakeholders than text \cite{Avison2003Book}. However, the visual syntax of software engineering languages has historically played a secondary role when comparing alternative visual notations for Software Engineering \cite{Moody2009TSE}. The confounding effect potentially played by language syntax is often \textbf{not} considered, when comparing languages. In his seminal paper on the \textit{``Physics'' of Notations} (PoN) \cite{Moody2009TSE}, Moody proposed a set of principles to support the evaluation, comparison, improvement and construction of visual notations for Software Engineering. His proposal focused on how to visually represent a set of constructs whose semantics had been previously defined. A core concept, adopted from \cite{Larkin1987CS}, is the notion of \textbf{cognitive effectiveness}, which can be defined as the \textbf{accuracy}, \textbf{speed}, and \textbf{ease} with which a representation can be processed by the human mind. 
\textbf{Semantic transparency}, together with the remaining 8 PoN principles, can lead to cognitive effectiveness. It is defined as \textit{``the extent to which the meaning of a symbol can be inferred from its appearance''} \cite{Moody2009TSE}.


Several studies were conducted on languages such as UML \cite{moody2008evaluating,el2015semantic}, BPMN \cite{Genon:2010:ACE:1964571.1964605,moody2011diagram}, KAOS \cite{matulevivcius2007visually} or \textit{i*} \cite{Moody2010REJ, Caire2013RE}, to identify improvement opportunities for those languages, by detecting problems concerning their concrete syntax and proposing solutions to mitigate them. Those studies focused on the stakeholders' ability to correctly recognise individual language symbols. However, software engineers use \textbf{models}, rather than their individual symbols, for communication.

In this paper, our objective is to compare the ability of stakeholders to \textit{understand} and \textit{review} social goal models using two concrete syntaxes: (i) the ``official'' \textit{i*} concrete syntax, and (ii) an 
alternative \textit{i*} concrete syntax, with an increased semantic transparency (that resulted from the series of experiments reported in \cite{Caire2013RE}). In particular, we performed a quasi-experiment to analyse the effect of changing the \textit{i*} concrete syntax, evaluating the 
semantic transparency impact on both the \textit{understandability} and the ability to \textit{review} \textit{i*} models, in the presence of a language key. Differently from previous studies, we perform our evaluation at the model level, rather than through isolated symbol recognition tasks.

A total of 57 novice participants (surrogates for stakeholders other than requirements engineers) performed understanding and reviewing tasks on \textit{i*} models. We measured the \textit{accuracy}, \textit{speed}, and \textit{ease} with which they accomplished their tasks. We found no evidence of improved accuracy or speed attributable to the 
alternative \textit{i*} concrete syntax, but found that working with this concrete syntax required significantly lower visual effort. 
This suggests that the usage of those symbols, in the context of models, and the presence of a language key, may have mitigated the \textit{i*} symbol recognition deficit consistently reported in previous works, to the point that it had no observable effect in the accuracy, or speed, with which our participants performed understanding and reviewing tasks.

Section \ref{sec:iStar} presents the two concrete syntaxes contrasted in this paper. Section \ref{sec:planning} reports the experiment planning, including goals, participants, experimental material, tasks, hypotheses, design, procedure, and analysis procedure. Section \ref{sec:execution} describes the experiment execution, with the preparation and deviations from the plan. Section \ref{sec:analysis} analyses the results, including descriptive statistics, dataset preparation, and the results of hypothesis testing. Section \ref{sec:discussion} discusses the results and reports threats to validity and inferences. Section \ref{sec:related_work} presents the related work. Finally, Section \ref{sec:conclusion} draws conclusions and points directions for future work. 

%% file: 3_new_iStar_Notation.tex
\section{i* standard and candidate notations}
\label{sec:iStar}

The \textit{i*} \cite{Yu1995} framework was designed for modelling and analysis of organisational environments and their information systems. \textit{Intentional actor} is the central concept of the approach. Actors are viewed as having intentional properties such as \textit{goals}, \textit{beliefs}, \textit{abilities} and \textit{commitments}. \textit{i*} has two main models: the Strategic Dependency (SD) and the Strategic Rationale (SR). The SD model describes the dependency relationships among the actors in an organisational context. An actor (the \textit{depender}) depends on another actor (the \textit{dependee}) to achieve goals and softgoals, to perform tasks and to obtain resources. The SR model focuses on modelling intentional elements and relationships internal to actors. 

\begin{figure}[!htpb] 
    \centering
    \centerline{\includegraphics[width=0.95\columnwidth]{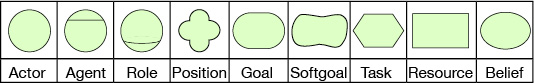}}
    \caption{Standard \textit{i*} symbol set}
    \label{fig:keyStandardNotation}
\end{figure}

Although well-known in the RE community, \textit{i*} is, as most other requirements languages, poorly understood by novice users \cite{Caire2013RE}. In this paper we explore the extent to which this problem can be mitigated by using 
an alternative concrete syntax for \textit{i*}, with an increased semantic transparency.
Research on visual languages design principles and evaluation has the potential for significantly improving the languages' adoption. One of the ways to improve the cognitive effectiveness of a notation is to increase the semantic transparency of its symbols. \textbf{Semantic transparency} defines the degree of association between the syntax (form) and semantic (content) of a symbol \cite{Moody2009TSE}. However, the \textit{i*} language concrete syntax, as described in the \textit{i*} Wiki\footnote{\url{http://istar.rwth-aachen.de/}}, has been shown to be \textbf{semantically opaque} \cite{Moody2010REJ}, as its symbols are abstract geometrical shapes (Fig. \ref{fig:keyStandardNotation}). In this paper, we will refer to this concrete syntax as ``standard'' \textit{i*}\footnote{\url{ http://istar.rwth-aachen.de/tiki-index.php?page=iStarQuickGuide}}.

The identification of this shortcoming in the concrete syntax of \textit{i*} led to the development of an alternative concrete syntax built upon the PoN principles, to make it more semantically transparent \cite{Moody2010REJ}. Later, Caire et al. reported a series of empirical evaluations involving 
the ``standard'' \textit{i*} and three alternative candidates:
\begin{inparaenum}[\itshape (i\upshape)]
    \item the symbols designed by experts following the PoN principles \cite{Moody2010REJ},
    \item the symbols more frequently designed by novices in the context of a symbolisation experiment, and
    \item the symbols more frequently chosen by subjects, among those designed by other novices \cite{genon2012towards}.
\end{inparaenum}

\begin{figure}[!htpb] 
    \centering
    \centerline{\includegraphics[width=0.95\columnwidth]{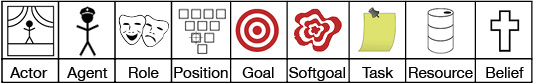}}
    \caption{ New \textit{i*} symbol set \cite{Caire2013RE}}
    \label{fig:keyNewNotation}
\end{figure}

Those four concrete syntaxes were then tested to determine which symbols were more frequently correctly identified in a blind interpretation experiment to evaluate the semantic transparency and cognitive load involved in recognising the symbols from the various \textit{i*} concrete syntaxes. 
The semantic transparency significantly increased symbol recognition and decreased interpretation errors \cite{Caire2013RE}. The symbol interpretation results were reported in such a way that it allowed us to choose the most frequently recognised symbol for each language construct (i.e., those with the highest semantic transparency coefficient), leading to a proposed \textit{i*} notation, referred in this paper as ``new'' \textit{i*} concrete syntax (Fig. \ref{fig:keyNewNotation}). 

We selected the best evaluated symbols in Caire et al.'s experiments for each \textit{i*} construct \cite{Caire2013RE}. As these symbols were selected independently from each other, they do not necessarily form a consistent set, in terms of the chosen visual metaphors, when compared to what an expert designer would be able to produce. Further research is required to study how an inconsistent set of symbols impacts the overall 
model understanding. Furthermore, these symbols might be difficult to draw by hand. In this paper, we are only covering 
model reading, 
with models 
built using an \textit{i*} editor. Thus, although important, the difficulty in drawing symbols by hand was not an issue in the present study.


%% file: 4_experiment.tex
\section{Experiment planning}
\label{sec:planning}

\subsection{Goals}
\label{sec:goals}

We describe our two research goals following the GQM research goals template \cite{Basili1988}. Our first goal (G1) is  to \textit{analyse} the effect of changing the \textit{i*} concrete syntax,
\textit{for the purpose of} evaluation, \textit{with respect to} its 
semantic transparency impact on the \textbf{understandability} of \textit{i*} SR models, \textit{from the viewpoint of} researchers,
\textit{in the context of} an experiment conducted with participants with limited or no experience with \textit{i*} at our University.
%
Our second  goal (G2) is to \textit{analyse} the effect of changing \textit{i*} concrete syntax, \textit{for the purpose of} evaluation, \textit{with respect to} its 
semantic transparency impact on the \textbf{review} of \textit{i*} SR models, \textit{from the viewpoint of} researchers, \textit{in the context of} an experiment conducted with the same participants.

Because we are comparing the effect of 
\textit{semantic transparency}
of two concrete syntaxes for the same abstract syntax, we can break down each goal into three sub-goals (G1.1, G1.2, G1.3, G2.1, G2.2, and G2.3), concerning the effect of those two concrete syntaxes, in terms of \textit{speed}, \textit{accuracy} and \textit{ease}. So, the refined goals can be obtained by replacing the term \textit{understandability} (or \textit{review}) with \textit{speed to understand}, \textit{effectiveness to understand} and \textit{ease to understand} (or \textit{speed to review}, \textit{effectiveness to review} and \textit{ease to review}). 

\subsection{Tasks}
\label{sec:tasks}
Before starting, each participant read and signed a letter of consent, adapted from \cite{Runeson2012Book}. Then, they saw a video with a small tutorial on \textit{i*}, covering all the model elements used in this evaluation. There were two versions of this video with exactly the same audio, but with the examples being presented in the standard \textit{i*} concrete syntax, or with the new \textit{i*} concrete syntax. Naturally, participants saw the video matching the concrete syntax they were about to use, in the evaluation.

Each participant in this study had to complete two tasks: \textit{understanding} an \textit{i*} SR model from a Goods Acquisition domain and \textit{reviewing} an \textit{i*} SR model from a Tolls System domain. In the \textit{understanding} task, the participant had to analyse a correct \textit{i*} SR model and answer a question about it. In the \textit{reviewing} task the participant had to analyse an incorrect \textit{i*} model and describe all the defects (s)he could identify. We deliberately introduced syntactic defects in the model. However, we have only informed the participants that their task was to find ``defects'', since describing explicitly the type of defects they should be looking for would have introduced a bias in the participants attention. This way, each participant was free to review the model using his best judgement, as a stakeholder new to \textit{i*} would.

In both tasks, the answers were recorded in audio, and we collected eye-tracking data while the participant was analysing each model. No eye-tracking feedback was visible to the participant, as this would be an unnecessary validity threat to the results. We also did not provided feedback on the extent to which participants were able to successfully complete the tasks, preventing possible contamination to subsequent tasks.

After each evaluation, participants filled in a NASA-TLX questionnaire \cite{Hart1988NASATLX,cao2009nasa} to collect feedback on his perceptions with respect to the task he had just performed. In the end, each participant provided some basic demographic information.

\subsection{Experimental material}
\label{sec:experimental_material}

\begin{figure*}[!htpb] 
	\centering
\begin{subfigure}[b]{0.9\columnwidth}
		\includegraphics[width=\columnwidth]{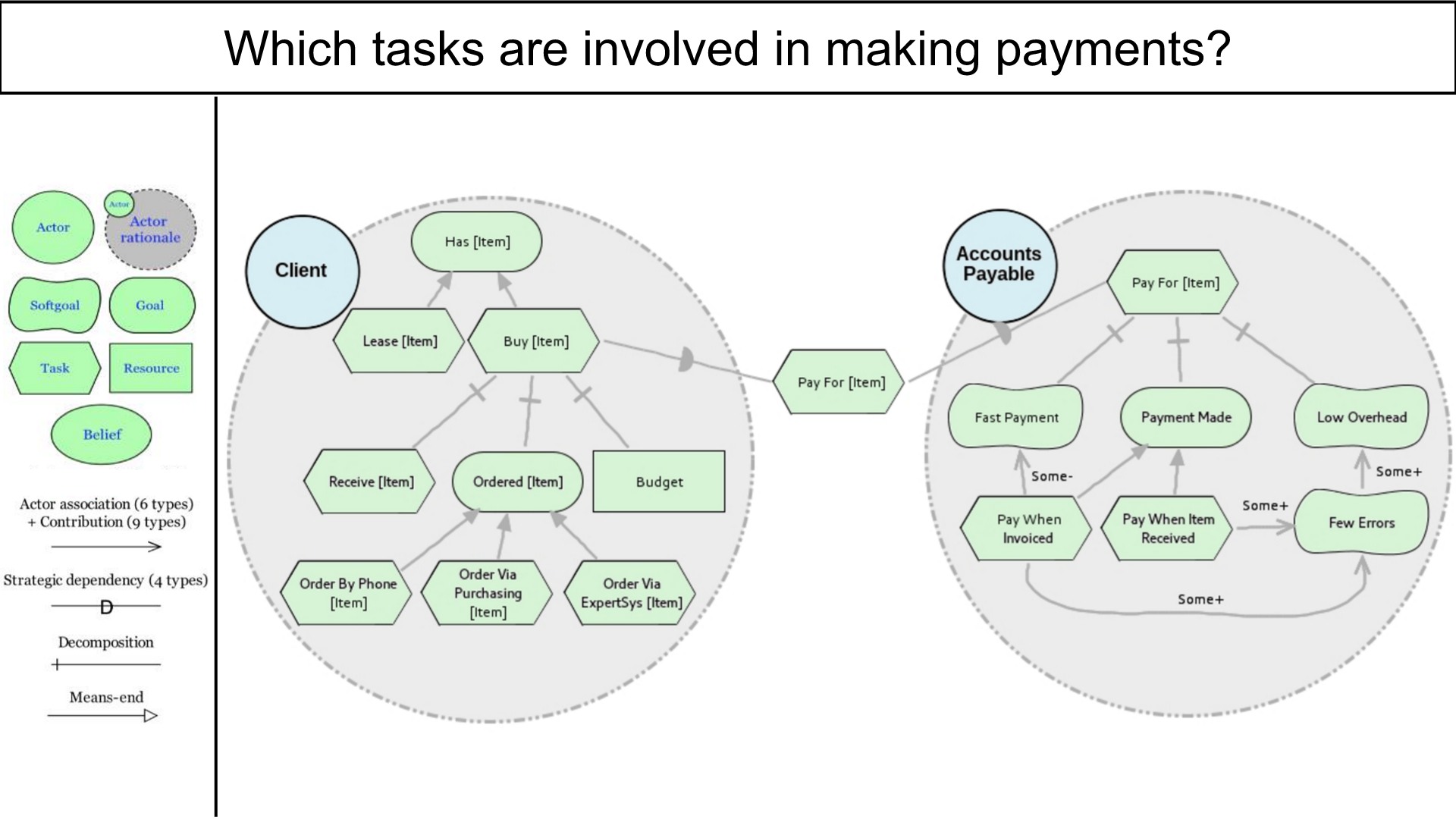}
		\caption{Comprehension task with the standard \textit{i*} concrete syntax}
		\label{fig:StandardUnderstandModel}
	\end{subfigure}
	~
	\begin{subfigure}[b]{0.9\columnwidth} 
		\includegraphics[width=\columnwidth]{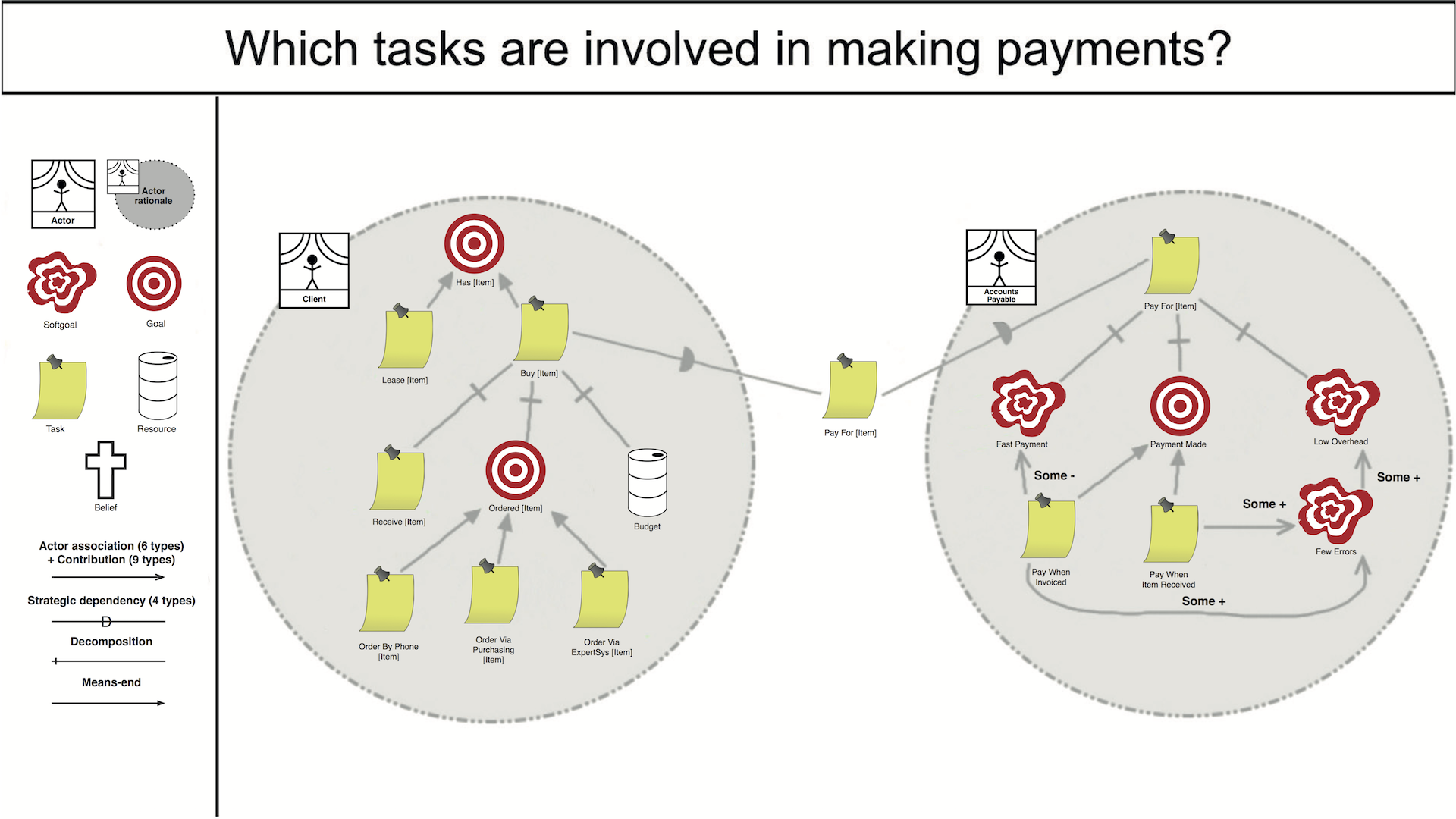}
		\caption{Comprehension task with the new \textit{i*} concrete syntax}
		\label{fig:NewUnderstandModel}
	\end{subfigure}

\vspace{0.30\baselineskip}

	\begin{subfigure}[b]{0.9\columnwidth}
		\includegraphics[width=\columnwidth]{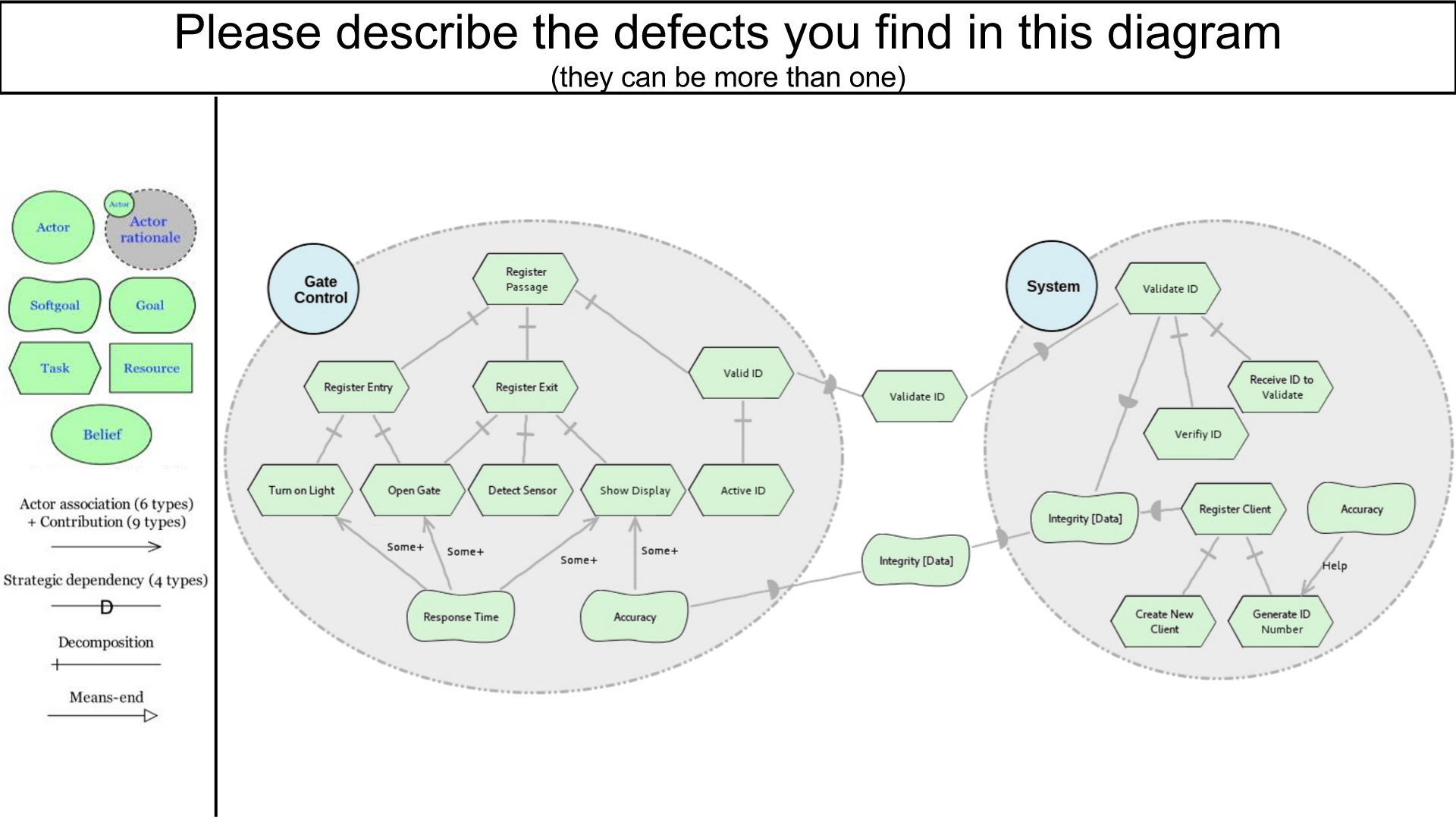}
		\caption{Review task with the standard \textit{i*} concrete syntax}
		\label{fig:StandardReviewModel}
	\end{subfigure}
	~
	\begin{subfigure}[b]{0.9\columnwidth}
		\includegraphics[width=\columnwidth]{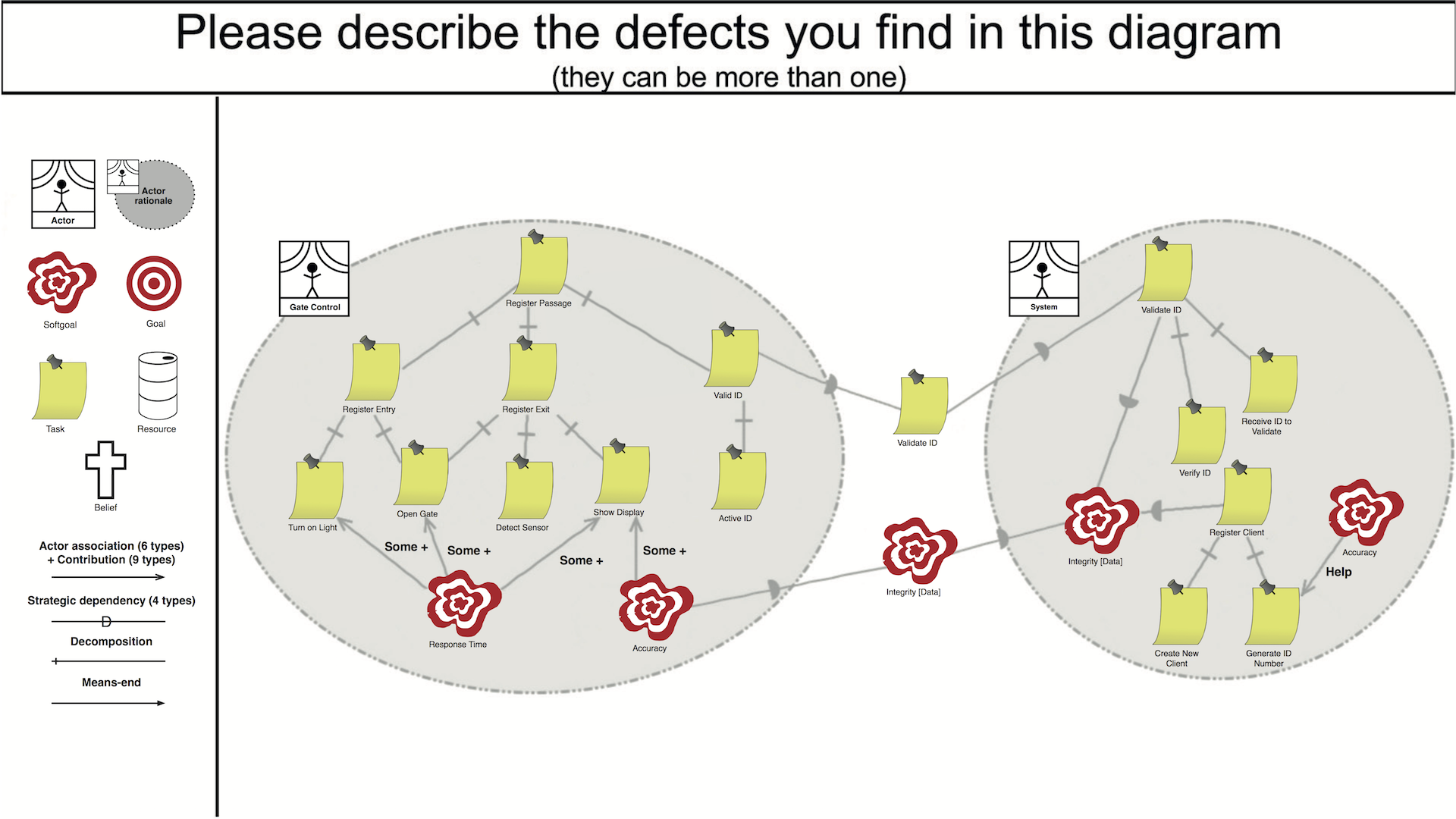}
		\caption{Review task with the new \textit{i*} concrete syntax}
		\label{fig:NewReviewModel}
	\end{subfigure}

	\caption{Understand and review tasks proposed to participants}
	\vspace{-12pt}

\end{figure*}

As previously mentioned, the experimental material for this evaluation included a participant consent letter, two video tutorials (one on the standard \textit{i*} concrete syntax and another on the \textit{new i*} concrete syntax), two versions of the \textit{i*} SR model for each of the two tasks (\textit{understanding} and \textit{reviewing}), a NASA-TLX questionnaire, and a demographic questionnaire. To contrast the two alternative concrete syntaxes for \textit{i*}, we prepared two versions of each \textit{i*} SR model, one with the standard \textit{i*} notation and the other with the new  \textit{i*} concrete syntax. The two versions of the model used for the understanding tasks are presented in Figs. \ref{fig:StandardUnderstandModel} and \ref{fig:NewUnderstandModel}. The two versions of the model for reviewing tasks are presented in Figs. \ref{fig:StandardReviewModel} and \ref{fig:NewReviewModel}. 
We were very conservative concerning readability. All the elements presented to participants, including textual labels, were comfortably readable in the 22 inch monitor used for conducting this experiment, so that readability would not be an issue. All figures share a common structure, with three Areas Of Interest (AOI): a \textbf{question} on top, a \textbf{language key} with the elements used in the 
model on the left side, and a \textbf{main area} with the model. For each task, we used a similar layout with both concrete syntaxes so that the only difference among them is the usage of a particular concrete syntax. For each task we annotated two sets of AOIs to analyse eye-tracking data. An AOI is classified as \textbf{relevant}, if it contains an element that belongs to the answer of the task, or \textbf{irrelevant}, otherwise. No textual descriptions of the scenarios under analysis were offered to the participants, so they had to answer our questions based only on the visual models.

Previous studies on \textit{i*} showed that its concrete syntax is semantically opaque, and that the alternative symbols used in this paper were easier to identify \cite{Caire2013RE}. The assumption was that these symbols would improve the understandability of the model by non-experts. However, those stakeholders are likely to examine \textit{i*} models in a document or within an \textit{i*} editor. In both cases, it is common to have a language key available: it is considered good practice to add a language key in a document, and the editor's toolbar will also serve as a language key, in practice. As such, the presence of a language key is aligned to common practice.

All the materials used in this evaluation, 
can be found in the paper's companion site\footnote{\url{https://sites.google.com/view/istarconcsyntaxexperiment/home}}. 

\subsection{Participants}
\label{sec:participants}

This evaluation was performed by 57 participants selected by convenience sampling. Most of them are students at different levels at our University. 
The main research question of this paper is whether improving the \textit{i*} 
concrete syntax has a real impact on its understandability by stakeholders other than Requirements Engineers. The latter may be specially trained to use this 
concrete syntax. Our target population is therefore non-experts, making our subjects better surrogates for this type of stakeholders than experienced RE practitioners. Students are often used as surrogates for practitioners in software engineering experiments \cite{Sjoberg2005TSE,Juristo2018} and have shown to be a valid option in those experiments  \cite{Host2000EMSE,Gorschek:2006}.

Of the total of participants, 27 were tested with the standard \textit{i*} concrete syntax, while the remaining 30 used the new concrete syntax. No participant was tested for both versions of the language, as a learning effect from one evaluation to the next could represent a confounding effect. Our goal was to have participants with a similar background testing both versions of the language. 
None of the participants who used the new concrete syntax participated in the evaluation of the standard concrete syntax, and \textit{vice-versa}. 

	

For each participant, we collected demographic data on previous \textit{experience} with \textit{i*}, \textit{age}, \textit{highest completed level of education}, \textit{current occupation} (student, researcher, or practitioner), \textit{field of studies}, \textit{gender}, \textit{nationality}, and \textit{usage of reading
devices} (glasses, or contact lenses).

Regarding previous \textit{experience} with \textit{i*}, in the group assigned to the standard concrete syntax, there was 1 participant who had used \textit{i*} in a professional context, 5 who learnt it in the context of a course, and the remaining 21
had no previous contact with \textit{i*}. For the new concrete syntax, 3 had learnt \textit{i*} in the context of a course and 27 did not know it.


Concerning participants \textit{age} distribution, the assumption of normality is \textbf{not} reasonable as shown by a Shapiro-Wilk test conducted on each of the participants groups ($p<0.001$, in both cases), and confirmed by the visual inspection of boxplots, Q-Q plots and kernel density plots, omitted here for the sake of brevity.
We then used the Welch \textit{t}-test to test if there was a statistically significant difference between the age distribution in the two groups. The Welch \textit{t}-test is robust when the group variances are unequal, and even if the sample sizes are unequal, as well as to departures from normality in the data. There was no statistically significant difference between the ages of participants using the standard \textit{i*} concrete syntax ($M=26.38$, $SD=5.933$) and those using the new concrete syntax ($M=24.63$, $SD=6.451$; $t(1)=53.792$, $p=.295$). 

With respect to the \textit{highest completed level of education}, all 
participants had some university level training. For those tested with the standard concrete syntax, 5 completed high school, 13 had BSc degrees, 8 had an MSc degree and 1 a PhD degree. For those participating in the new concrete syntax evaluation, 7 completed high school, 18 had a BSc degree, 4 an MSc degree, and 1 a PhD degree. Concerning current occupation, the standard concrete syntax had 1 researcher, 3 practitioners, and the remainder were students; the new concrete syntax, had 1 researcher, 1 practitioner and the remainder were students.

Regarding \textit{nationality}, 21 Portuguese, 4 Brazilians, 1 Croatian and 1 Spaniard used the standard concrete syntax, and 29 Portuguese and 1 Brazilian used the new concrete syntax. 

Concerning the \textit{field of studies}, for the standard concrete syntax, 25 were computer scientists, and 2 were industrial engineering. For the new concrete syntax, we had 22 computer scientists, 2 industrial engineers, 2 architects, 1 mechanical engineer, 1 manager, 1 civil engineer and 1 lawyer. For each concrete syntax, there were 4 female participants. The remainder were male. In terms of the usage of \textit{reading devices}, 3 participants using the standard concrete syntax and 1 using the new concrete syntax had contact lenses while, in each group 4 and 8, respectively, wore glasses.

\subsection{Hypotheses, parameters and variables}
\label{sec:hypotheses}

For each of the two high level goals, we define the null ($H_0$) and alternative hypotheses ($H_1$). 

\begin{framed} 
\small
$H_{0Understand}$: Changing from a semantically opaque concrete syntax (standard \textit{i*}) to a more semantically transparent one (new \textit{i*}) does not influence \textit{i*} SR models \textit{understandability}.

$H_{1Understand}$: Changing from a semantically opaque concrete syntax (standard \textit{i*}) to a more semantically transparent one (new \textit{i*}) influences \textit{i*} SR models \textit{understandability}.
\end{framed}



This hypothesis is further refined to cope with \textit{accuracy}, \textit{speed} and \textit{effort}. For example:

\begin{framed} 
\small
$H_{0UnderstandAcc}$: Changing from a semantically opaque concrete syntax (standard \textit{i*}) to a more semantically transparent one (new \textit{i*}) does not influence \textit{i*} SR models \textit{understanding accuracy}.

$H_{1UnderstandAcc}$: Changing from a semantically opaque concrete syntax (standard \textit{i*}) to a more semantically transparent one (new \textit{i*}) influences \textit{i*} SR models \textit{understanding accuracy}.
\end{framed}



And similarly for speed and ease of \textit{understanding}. 
%
%
%
%
%
%
%
We follow the same approach and refine the \textit{null} and the \textit{alternative} hypotheses in the case of \textit{review} into 3 sub-hypotheses, corresponding to \textit{accuracy}, \textit{speed} and \textit{effort}. The independent variable is the \textit{concrete syntax}, which may be \textit{standard}, or \textit{new}. The dependent variables are the same for both hypotheses, as well as their corresponding refined sub-hypotheses.

\textbf{Assessing accuracy.} The accuracy achieved by our participants is assessed by their responses with respect to their \textit{precision}, and \textit{recall}, using the following metrics:

\begin{itemize}
	\item \textit{precision} -- the fraction of model elements retrieved by participants (for the first hypothesis) or of defects (for the second hypothesis) which are relevant.
	\item \textit{recall} -- the fraction of relevant model elements (or of relevant defects) retrieved by participants, over the total number of model elements (or potential defects) retrieved.
	\item \textit{F-measure} -- a measure that combines precision and recall, computed as $\frac{2 * (Precision * Recall)}{(Precision + Recall)}$; this measure provides an harmonic mean of precision and recall.
\end{itemize}

Higher values of \textit{Precision}, \textit{Recall}, and the \textit{F-measure}, support the claim of a better accuracy.

\textbf{Assessing speed.} The speed achieved by our participants is assessed by several time-related indicators.
We are interested not only in the overall response time, but also on the time it takes participants to provide valid answers. We will assess 
\textit{speed} using the following metrics:

\begin{itemize}
	\item \textit{Duration} -- the time taken by the participants to complete the task.
	\item \textit{FirstDet} -- First Detection; the time taken to accurately report the \textit{first} response element; for the understanding task, this is the time for correctly reporting the first element that answers the question enunciated in the task; for the reviewing task, this is the time taken to report the first seeded defect in the model. If a participant does not correctly report at least one element, this metric will be treated as a missing value and removed from all further analysis procedures.
	\item \textit{LastDet} -- Last Detection; the time taken to accurately report the \textit{last} response element; this is the dual for the \textit{FirstDet} metric.
\end{itemize}

Lower values of these metrics support the claims of superiority of the corresponding concrete syntax with respect to its cognitive effectiveness in terms of improving the speed with which the models are understood and reviewed. While the overall \textit{duration} addresses the time spent in the task, the other two metrics provide a detailed picture of the moment when the participant starts and ends providing valid feedback. 

\textbf{Assessing ease.} The ease with which participants conduct their tasks is assessed by effort measures.
Although time measures (as those we used for speed) are often used as proxies for effort, in the context of the \textit{``Physics'' of Notations} these are better matches for the speed component, which is likely to strongly correlate to ease. Instead, we focus our assessment in two information sources: the physical (\textit{visual}) effort involved in exploring the model and the \textit{perception of effort} reported by participants. The former is addressed with eye-tracking measurements, while the latter is assessed through a NASA-TLX questionnaire. 
We will consider the following metrics:

\begin{itemize}
	\item \textit{FixRel} -- Fixation Rate on Relevant elements; the fraction of number of fixations in an given AOI over the total number of fixations in the AOG (Area of Glance). A fixation is a stabilisation of the eye on a part of the stimulus for a period of time between 200 and 300 ms.
	\item \textit{FixIrrel} -- Fixation Rate on Irrelevant elements; the fraction of number of fixations in an given AOI over the total number of fixations in the AOG.
	\item \textit{AvDurFixRel} -- Average Duration of Relevant Fixation; the fraction of total duration of fixations for relevant AOIs over the number of elements of the relevant AOIs.
	\item \textit{AvDurFixIrrel} -- Average Duration of Irrelevant Fixation; the fraction of total duration of fixations for irrelevant AOIs over the number of elements of the irrelevant AOIs. 
    \item \textit{TotSac} -- total number of saccades while performing the task. A saccade is a sudden and quick eye-movement lasting between 40 to 50 ms.
    \item \textit{Sac2Key} -- number of saccades to the key AOI.
	\item \textit{NASA-TLX score} -- overall weighted score resulting from the application of the TLX questionnaire, covering perceived mental, physical and temporal demand, performance, effort and frustration for 
	performing a task.
\end{itemize}


A higher number and duration of fixations is associated with a higher visual attention in a given set of AOIs (in this case, relevant \textit{vs.} irrelevant model elements) 
\cite{Shaffer2015,sharafi2015systematic,poole2006eye}.
For understating tasks, a higher Fixation Rate indicates higher efficiency associated with less effort to find the relevant AOIs \cite{poole2006eye,porras2010empirical,sharif2010eye,sharif2012eye, de2014taupe}.
As for reviewing tasks, a higher ratio indicates more visual effort to find defects \cite{sharif2010eye,sharif2013empirical}. Regarding the Average Fixation Duration, a higher value indicates more time and attention devoted to 
AOIs \cite{sharafi2015systematic, porras2010empirical,cagiltay2013performing}, some state this ratio is correlated with cognitive processes \cite{duchowski2007eye,goldberg1999computer}.
A higher number of saccades can be associated with a higher visual effort, meaning 
the participant may be somewhat ``lost'' in the model, making a more erratic model navigation  \cite{de2014taupe,fritz2014using,goldberg1999computer, sharafi2015systematic}. A higher number of saccades to the key can also be associated with difficulties with the concrete syntax.  Concerning the NASA-TLX score, higher scores are associated with a higher perceived effort by the participants \cite{cao2009nasa,fritz2014using}. Both for the eye-tracking and the NASA-TLX metrics, lower complexity will correspond to higher \textit{ease} in performing the tasks.

\subsection{Design}
\label{sec:design}
Data collection was performed in two different moments, one for each concrete syntax. Due to participants availability constraints, most of those using the standard \textit{i*} concrete syntax performed only either the understanding or the review task. 
Three of them performed both. As for the participants using the new \textit{i*} concrete syntax, they all performed both 
tasks.
To reduce learning effects, for those performing 2 tasks, the relative order of those tasks changed from one participant to the next. Each participant worked only with one of the concrete syntaxes for \textit{i*}. We balanced the number of times each task was performed before, or after the other task.

The sequence experienced by each participant is illustrated in Table \ref{tab:design}, where each line represents a set of participants that followed a particular sequence of activities. T\# refers to the task number and  Back to the background questionnaire (demographic data). The tasks are encoded: concerning the first character, \textbf{U} stands for \textbf{u}nderstand, while \textbf{R} stands for \textbf{r}eview; the second character 
represents the particular concrete syntax used by that participant, where \textbf{S} stands for 
\textbf{S}tandard \textit{i*} concrete syntax and \textbf{N} stands for \textbf{N}ew \textit{i*} concrete syntax. There was no pre-defined sequence for ordering participants.  

\begin{table}[ht]
	\centering
	\footnotesize

	\caption{Experimental design}
	\label{tab:design}
	\begin{tabular}{@{}llllllll@{}}
		\toprule
		\#Participants &  Letter    & Tutorial   & T1  & TLX & T2  & TLX & Back\\ \midrule
		13  & \checkmark & \checkmark & US & \checkmark &    &    & \checkmark \\
		11           & \checkmark & \checkmark & RS & \checkmark &  &  & \checkmark \\
		2           & \checkmark & \checkmark & US & \checkmark & RS & \checkmark & \checkmark \\
		1           & \checkmark & \checkmark & RS & \checkmark & US & \checkmark & \checkmark \\ \midrule
		15           & \checkmark & \checkmark & UN & \checkmark & RN &  \checkmark & \checkmark \\
		15           & \checkmark & \checkmark & RN & \checkmark & UN & \checkmark & \checkmark \\\bottomrule
	\end{tabular}
\end{table}

The statistical analysis performed (Welch \textit{t}-test) is robust concerning the different sample sizes, that is, a different number of participants performing each sequence.

\subsection{Procedure}
\label{sec:procedure}

We prepared the lab setting so that all participants could have similar conditions. There was only one participant in each evaluation session. We informed him that the tasks consisted in watching a short tutorial on a requirements language, analysing requirements expressed in that language, and answering questions about those requirements. We further informed the participants that we would be recording their voice, the contents of the screen, and tracking their eyes movements while they were analysing the requirements and (orally) answering questions about them.
Finally, we explained they could quit at any moment, if they so desired. They then read the \textit{Participant consent letter} and gave their free and informed consent to participate in the study.

We helped the participant sit comfortably so that his eyes would be around 50 cm away from the screen. The eye-tracker was placed below the screen, without blocking it. We adjusted the eye-tracker's angle to cope with physical differences among the participants (the eye-tracker must point towards the subject's eyes, so the participant's height determines the ideal eye-tracker angle). 
The participant put on the headphones (equipped with a microphone), and the session started.

We asked each participant to watch a video tutorial of 7 minutes and 15 seconds, explaining the elements of an \textit{i*} model. The tutorial includes the construction of a correct model, similar to those used in the experiment, and an audio description of both the modelling elements, as they are being introduced, and their role in the model under construction. The modelling elements were described using the exact phrases and explanations present in the \textit{i*} wiki. At the end of this tutorial, we calibrated the eye-tracker, 
and started the evaluation session.
Each participant was asked to perform a sequence of two 
tasks. Each task consisted in either understanding or reviewing an \textit{i*} model, and then answering the NASA-TLX questionnaire concerning the effort on that task. This was repeated for each task. The task (and corresponding model) sequence varied from one participant to the next (discussed in Section \ref{sec:design}). Finally, each participant answered a short questionnaire about demographic information. For each session, we recorded a video with the contents of the screen, synchronised with the voice of the subject during the whole session. We also recorded the  NASA-TLX sets of answers, one for each task, and the answers to the demographic questionnaire.

\subsection{Analysis procedure}
\label{sec:analysis_procedure}
We collected descriptive statistics on our variables, namely the \textit{mean}, \textit{standard deviation}, \textit{skewness} and \textit{kurtosis}, to get an overview of their distribution. This was complemented with kernel density plots to help with the visual analysis of those distributions. Kernel density plots provide a more detailed picture of a distribution, when compared to boxplots, and are a better fit for comparing distributions in Software Engineering experimentation. This visual analysis was then complemented with Welch \textit{t}-tests, which provide an 
alternative to the t-test, as they can robustly handle non-normal distributions, with different sample sizes and 
variances. Section \ref{sec:analysis} shows that the vast majority share these properties. A detailed discussion on the benefits of using kernel density plots \textit{vs.} box plots, and using Welch \textit{t}-test for comparing distributions in a robust way 
(as opposed to two samples t-test, or a non-parametric alternative to it, such as the Mann-Whitney U test) is in \cite{Kitchenham2016EmSE}.

%% file: 5_execution.tex
\section{Execution}
\label{sec:execution}

\subsection{Preparation}
\label{sec:preparation}

The data collection was carried out with a laptop connected to an external 22 inch, wide screen, full HD monitor, an EyeTribe eye-tracker\footnote{\url{http://www.theeyetribe.com/}}, a set of headphones 
with a microphone, and an external mouse and keyboard. The experimenter controlled the session on the laptop, while the participant used the eye-tracker, 
headphones 
and microphone to perform the models' analysis, viewing the tasks in the external monitor. Each participant started by reading a consent information letter, then watched the video tutorial on the \textit{i*} framework. That was the only source of information on \textit{i*} the participant would have for the duration of the experiment, other than an \textit{i*} language key (see Fig. \ref{fig:NewReviewModel}).

Finally, we recorded the audio and video of the whole section, so that the answers were collected with a \textit{think aloud} approach. We proceeded with the calibration of the eye-tracker, which consists of having the participant following with her gaze a target as it moves and fixates in predetermined screen coordinates. We used the EyeTribe calibration application, only accepting \textit{good} or \textit{excellent} calibrations (top levels of a 5 points ordinal scale) to proceed to the actual data collection. 

\subsection{Deviations}
\label{sec:deviations}

During the standard \textit{i*} concrete syntax experiment, we observed a technical problem with the software for audio capturing, leading to the exclusion of a total of 6 cases (4 Rev. + 2 Und.). This can be perceived in Table \ref{tab:DescriptiveStats}, on sector for accuracy metrics where the number of participants is 12 for both tasks (instead of 14 and 16).
Another situation in this same experiment, where we could not determine when the participant started viewing each of the models led to the partial exclusion of 1 case (Und.).
In addition, a technical problem with the eye-tracker device led to the exclusion of 2 cases for the understanding task, one for each concrete syntax. This can be observed in Table \ref{tab:DescriptiveStats} on the sectors for speed and visual effort metrics, where the total number for the understanding task is 10 and 29 instead of 12 and 30, respectively.

%% file: 6_analysis.tex
\section{Analysis}
\label{sec:analysis}

\subsection{Descriptive statistics}
\label{sec:statistics}
Table \ref{tab:DescriptiveStats} presents the descriptive statistics for the metrics collected in our data analysis, 
(introduced in section \ref{sec:hypotheses}). For each metric we present 4 lines in the table. The first 2 
refer to the understanding task, while the other 2 
refer to the reviewing task. In the \textit{Syntax} column we specify which of the syntaxes we are considering (\textit{Stand.} represents the standard \textit{i*} concrete syntax, while \textit{New} represents the new concrete syntax.
We further present the mean, standard deviation, skewness, kurtosis, and the \textit{p-value} for the Shapiro-Wilk normality test. The number of participants is not always the same for all metrics (as per section \ref{sec:deviations}), and 
missing values were excluded from the analysis 
due to 
anomalies in the data collection process (e.g., 
a situation 
when no valid elements were detected, it made no sense to compute the corresponding metric; this is particularly noticeable in the review task).

The metrics are visually grouped to reflect the three components of cognitive effectiveness: \textit{accuracy}, \textit{speed} and \textit{ease}. One of the most 
noteworthy features of our data set is that the shape of the distributions concerning variables related to \textit{accuracy} and \textit{speed} suggests that, in general, normality is \textbf{not} a reasonable assumption (\textit{p-value} $<$ 0.05). 
Several of the metrics concerning \textit{ease} do have a distribution suggesting normality is a reasonable assumption (\textit{p-value} $\ge$ 0.05). The variance of the distributions is not similar, for several of these variables. The visual inspection of boxplot diagrams, Q-Q plots and kernel density plots (ommitted here for the sake of brevity) further reinforced our assessment concerning data normality.

\begin{table}[htb]
	\centering
		\vspace{-8pt}
	\footnotesize
	\caption{Descriptive statistics}
	\label{tab:DescriptiveStats}
	\begin{tabular}{|l|p{0.4cm}p{0.7cm}p{0.2cm}p{0.65cm}p{0.5cm}p{0.7cm}p{0.78cm}p{0.6cm}|}
		\hline
		&\textbf{Task} & \textbf{Syntax} & \textbf{\#} & \textbf{Mean} & \textbf{S.D.} & \textbf{Skew} & \textbf{Kurt} & \textbf{S-W} \\ \hline \hline
		
		\multirow{4}{1mm}{\begin{sideways}Prec.\end{sideways}} & \multirow{2}{*}{Und.} & Stand. & 12 & \multicolumn{1}{r}{.653} & \multicolumn{1}{r}{.325} & \multicolumn{1}{r}{-.638} & \multicolumn{1}{r}{-.115} & \multicolumn{1}{r|}{.146} \\
		& & New & 30 & \multicolumn{1}{r}{.525} & \multicolumn{1}{r}{.291} & \multicolumn{1}{r}{.158} & \multicolumn{1}{r}{-1.075} & \multicolumn{1}{r|}{.048} \\ \cline{2-9}
		
		& \multirow{2}{*}{Rev.} & Stand. & 12 & \multicolumn{1}{r}{.089} & \multicolumn{1}{r}{.215} & \multicolumn{1}{r}{2.363} & \multicolumn{1}{r}{4.881} & \multicolumn{1}{r|}{.000} \\ 
		& & New & 30 & \multicolumn{1}{r}{.131} & \multicolumn{1}{r}{.271} & \multicolumn{1}{r}{2.135} & \multicolumn{1}{r}{3.955} & \multicolumn{1}{r|}{.000} \\ \hline
		
		\multirow{4}{1mm}{\begin{sideways}Recall\end{sideways}} & \multirow{2}{*}{Und.} & Stand. & 12 & \multicolumn{1}{r}{.722} & \multicolumn{1}{r}{.446} & \multicolumn{1}{r}{-1.181} & \multicolumn{1}{r}{-.584} & \multicolumn{1}{r|}{.000} \\
		& & New & 30 & \multicolumn{1}{r}{.678} & \multicolumn{1}{r}{.309} & \multicolumn{1}{r}{-.347} & \multicolumn{1}{r}{-1.172} & \multicolumn{1}{r|}{.000} \\ \cline{2-9}
		
		& \multirow{2}{*}{Rev.} & Stand. & 12 & \multicolumn{1}{r}{.048} & \multicolumn{1}{r}{.111} & \multicolumn{1}{r}{2.055} & \multicolumn{1}{r}{2.640} & \multicolumn{1}{r|}{.000} \\ 
		& & New & 30 & \multicolumn{1}{r}{.067} & \multicolumn{1}{r}{.190} & \multicolumn{1}{r}{4.390} & \multicolumn{1}{r}{21.296} & \multicolumn{1}{r|}{.000} \\ \hline
		
		\multirow{4}{1mm}{\begin{sideways}F-Meas.\end{sideways}} & \multirow{2}{*}{Und.} & Stand. & 12 & \multicolumn{1}{r}{.615} & \multicolumn{1}{r}{.415} & \multicolumn{1}{r}{-.713} & \multicolumn{1}{r}{-1.241} & \multicolumn{1}{r|}{.011} \\
		& & New & 30 & \multicolumn{1}{r}{.573} & \multicolumn{1}{r}{.280} & \multicolumn{1}{r}{-.175} & \multicolumn{1}{r}{-1.216} & \multicolumn{1}{r|}{.026} \\ \cline{2-9}
		
		&\multirow{2}{*}{Rev.} & Stand. & 12 & \multicolumn{1}{r}{.061} & \multicolumn{1}{r}{.143} & \multicolumn{1}{r}{2.100} & \multicolumn{1}{r}{2.974} & \multicolumn{1}{r|}{.000} \\
		& & New & 30 & \multicolumn{1}{r}{.079} & \multicolumn{1}{r}{.189} & \multicolumn{1}{r}{3.515} & \multicolumn{1}{r}{14.577} & \multicolumn{1}{r|}{.000} \\ \hline \hline
		
		\multirow{4}{1mm}{\begin{sideways}Duration\end{sideways}} & \multirow{2}{*}{Und.} & Stand. & 10 & \multicolumn{1}{r}{131.9} & \multicolumn{1}{r}{90.0} & \multicolumn{1}{r}{2.291} & \multicolumn{1}{r}{5.850} & \multicolumn{1}{r|}{.001} \\
		& & New & 29 & \multicolumn{1}{r}{163.8} & \multicolumn{1}{r}{111.1} & \multicolumn{1}{r}{1.572} & \multicolumn{1}{r}{2.698} & \multicolumn{1}{r|}{.001} \\ \cline{2-9}
		&\multirow{2}{*}{Rev.} & Stand. & 12 & \multicolumn{1}{r}{255.3} & \multicolumn{1}{r}{179.4} & \multicolumn{1}{r}{1.755} & \multicolumn{1}{r}{3.815} & \multicolumn{1}{r|}{.014} \\
		& & New & 30 & \multicolumn{1}{r}{263.9} & \multicolumn{1}{r}{143.9} & \multicolumn{1}{r}{.734} & \multicolumn{1}{r}{-.177} & \multicolumn{1}{r|}{.059} \\ \hline
		
		\multirow{4}{1mm}{\begin{sideways}FirstDet\end{sideways}} & \multirow{2}{*}{Und.} & Stand. & 10 & \multicolumn{1}{r}{120.2} & \multicolumn{1}{r}{103.0} & \multicolumn{1}{r}{2.138} & \multicolumn{1}{r}{5.340} & \multicolumn{1}{r|}{.007} \\
		& & New & 28 & \multicolumn{1}{r}{106.1} & \multicolumn{1}{r}{75.9} & \multicolumn{1}{r}{2.015} & \multicolumn{1}{r}{4.969} & \multicolumn{1}{r|}{.000} \\ \cline{2-9}
		
		&\multirow{2}{*}{Rev.} & Stand. & 2 & \multicolumn{1}{r}{174.5} & \multicolumn{1}{r}{137.9} & \multicolumn{1}{r}{-} & \multicolumn{1}{r}{-} & \multicolumn{1}{r|}{-} \\
		& & New & 7 & \multicolumn{1}{r}{192.0} & \multicolumn{1}{r}{234.0} & \multicolumn{1}{r}{2.327} & \multicolumn{1}{r}{5.633} & \multicolumn{1}{r|}{.002} \\ \hline
		
		\multirow{4}{1mm}{\begin{sideways}LastDet\end{sideways}} & \multirow{2}{*}{Und.} & Stand. & 10 & \multicolumn{1}{r}{126.3} & \multicolumn{1}{r}{104.2} & \multicolumn{1}{r}{2.037} & \multicolumn{1}{r}{4.941} & \multicolumn{1}{r|}{.013} \\
		& & New & 28 & \multicolumn{1}{r}{113.6} & \multicolumn{1}{r}{78.8} & \multicolumn{1}{r}{1.906} & \multicolumn{1}{r}{4.385} & \multicolumn{1}{r|}{.000} \\ \cline{2-9}
		
		&\multirow{2}{*}{Rev.} & Stand. & 2 & \multicolumn{1}{r}{217.0} & \multicolumn{1}{r}{89.1} & \multicolumn{1}{r}{-} & \multicolumn{1}{r}{-} & \multicolumn{1}{r|}{-} \\
		& & New & 7 & \multicolumn{1}{r}{204.6} & \multicolumn{1}{r}{227.0} & \multicolumn{1}{r}{2.373} & \multicolumn{1}{r}{5.860} & \multicolumn{1}{r|}{.001} \\ \hline\hline
		
		\multirow{4}{1mm}{\begin{sideways}RelFix\end{sideways}} & \multirow{2}{*}{Und.} & Stand. & 10 & \multicolumn{1}{r}{.127} & \multicolumn{1}{r}{.158} & \multicolumn{1}{r}{.701} & \multicolumn{1}{r}{-1.614} & \multicolumn{1}{r|}{.004} \\
		& & New & 29 & \multicolumn{1}{r}{.135} & \multicolumn{1}{r}{.095} & \multicolumn{1}{r}{1.081} & \multicolumn{1}{r}{.470} & \multicolumn{1}{r|}{.005} \\ \cline{2-9}
		& \multirow{2}{*}{Rev.} & Stand. & 12 & \multicolumn{1}{r}{.086} & \multicolumn{1}{r}{.049} & \multicolumn{1}{r}{.257} & \multicolumn{1}{r}{.073} & \multicolumn{1}{r|}{.984} \\ 
		& & New & 30 & \multicolumn{1}{r}{.026} & \multicolumn{1}{r}{.031} & \multicolumn{1}{r}{1.741} & \multicolumn{1}{r}{3.205} & \multicolumn{1}{r|}{.001} \\ 
		\hline
		
		\multirow{4}{1mm}{\begin{sideways}IrrelFix\end{sideways}} & \multirow{2}{*}{Und.} & Stand. & 10 & \multicolumn{1}{r}{.293} & \multicolumn{1}{r}{.201} & \multicolumn{1}{r}{.212} & \multicolumn{1}{r}{-1.305} & \multicolumn{1}{r|}{.373} \\
		& & New & 29 & \multicolumn{1}{r}{.264} & \multicolumn{1}{r}{.259} & \multicolumn{1}{r}{.114} & \multicolumn{1}{r}{.-.898} & \multicolumn{1}{r|}{.559} \\ \cline{2-9}
		& \multirow{2}{*}{Rev.} & Stand. & 12 & \multicolumn{1}{r}{.282} & \multicolumn{1}{r}{.100} & \multicolumn{1}{r}{-1.019} & \multicolumn{1}{r}{1.344} & \multicolumn{1}{r|}{.376} \\ 
		& & New & 30 & \multicolumn{1}{r}{.370} & \multicolumn{1}{r}{.112} & \multicolumn{1}{r}{-.232} & \multicolumn{1}{r}{-.337} & \multicolumn{1}{r|}{.573} \\ 
		\hline
		\multirow{4}{1mm}{\begin{sideways}AvRelDur\end{sideways}} & \multirow{2}{*}{Und.} & Stand. & 10 & \multicolumn{1}{r}{174.0} & \multicolumn{1}{r}{213.2} & \multicolumn{1}{r}{.671} & \multicolumn{1}{r}{-1.464} & \multicolumn{1}{r|}{.008} \\
		& & New & 29 & \multicolumn{1}{r}{327.7} & \multicolumn{1}{r}{153.7} & \multicolumn{1}{r}{.322} & \multicolumn{1}{r}{-.392} & \multicolumn{1}{r|}{.832} \\ \cline{2-9}
		& \multirow{2}{*}{Rev.} & Stand. & 11 & \multicolumn{1}{r}{323.2} & \multicolumn{1}{r}{134.0} & \multicolumn{1}{r}{.800} & \multicolumn{1}{r}{.056} & \multicolumn{1}{r|}{.184} \\ 
		& & New & 22 & \multicolumn{1}{r}{274.7} & \multicolumn{1}{r}{205.2} & \multicolumn{1}{r}{1.775} & \multicolumn{1}{r}{4.478} & \multicolumn{1}{r|}{.003} \\
		
		\hline
		\multirow{4}{1mm}{\begin{sideways}AvIrrelDur\end{sideways}} & \multirow{2}{*}{Und.} & Stand. & 10 & \multicolumn{1}{r}{305.1} & \multicolumn{1}{r}{98.0} & \multicolumn{1}{r}{.476} & \multicolumn{1}{r}{-.432} & \multicolumn{1}{r|}{.724} \\
		& & New & 29 & \multicolumn{1}{r}{289.2} & \multicolumn{1}{r}{114.6} & \multicolumn{1}{r}{.385} & \multicolumn{1}{r}{-.732} & \multicolumn{1}{r|}{.412} \\ \cline{2-9}
		& \multirow{2}{*}{Rev.} & Stand. & 12 & \multicolumn{1}{r}{238.0} & \multicolumn{1}{r}{57.3} & \multicolumn{1}{r}{.127} & \multicolumn{1}{r}{-1.976} & \multicolumn{1}{r|}{.039} \\ 
		& & New & 30 & \multicolumn{1}{r}{292.7} & \multicolumn{1}{r}{103.4} & \multicolumn{1}{r}{1.795} & \multicolumn{1}{r}{5.056} & \multicolumn{1}{r|}{.001} \\ 
		\hline
		
		\multirow{4}{1mm}{\begin{sideways}TotSac\end{sideways}} & \multirow{2}{*}{Und.} & Stand. & 10 & \multicolumn{1}{r}{47.7} & \multicolumn{1}{r}{12.6} & \multicolumn{1}{r}{.139} & \multicolumn{1}{r}{-1.795} & \multicolumn{1}{r|}{.035} \\
		& & New & 29 & \multicolumn{1}{r}{41.7} & \multicolumn{1}{r}{21.5} & \multicolumn{1}{r}{.266} & \multicolumn{1}{r}{-.587} & \multicolumn{1}{r|}{.459} \\ \cline{2-9}
		& \multirow{2}{*}{Rev.} & Stand. & 12 & \multicolumn{1}{r}{460.0} & \multicolumn{1}{r}{324.9} & \multicolumn{1}{r}{1.653} & \multicolumn{1}{r}{3.314} & \multicolumn{1}{r|}{.027} \\ 
		& & New & 30 & \multicolumn{1}{r}{458.8} & \multicolumn{1}{r}{256.5} & \multicolumn{1}{r}{.568} & \multicolumn{1}{r}{-.557} & \multicolumn{1}{r|}{.128} \\ \hline
		
		\multirow{4}{1mm}{\begin{sideways}Sac2Key\end{sideways}} & \multirow{2}{*}{Und.} & Stand. & 10 & \multicolumn{1}{r}{165.4} & \multicolumn{1}{r}{28.7} & \multicolumn{1}{r}{-.959} & \multicolumn{1}{r}{.057} & \multicolumn{1}{r|}{.174} \\
		& & New & 29 & \multicolumn{1}{r}{117.6} & \multicolumn{1}{r}{55.8} & \multicolumn{1}{r}{-.910} & \multicolumn{1}{r}{-.006} & \multicolumn{1}{r|}{.008} \\ \cline{2-9}
		& \multirow{2}{*}{Rev.} & Stand. & 12 & \multicolumn{1}{r}{83.75} & \multicolumn{1}{r}{51.9} & \multicolumn{1}{r}{-.142} & \multicolumn{1}{r}{-.753} & \multicolumn{1}{r|}{.795} \\ 
		& & New & 30 & \multicolumn{1}{r}{141.3} & \multicolumn{1}{r}{52.0} & \multicolumn{1}{r}{-1.587} & \multicolumn{1}{r}{2.724} & \multicolumn{1}{r|}{.000} \\ \hline
		
		
		\multirow{4}{1mm}{\begin{sideways}TLX\end{sideways}} & \multirow{2}{*}{Und.} & Stand. & 16 & \multicolumn{1}{r}{47.7} & \multicolumn{1}{r}{12.6} & \multicolumn{1}{r}{.139} & \multicolumn{1}{r}{-1.795} & \multicolumn{1}{r|}{.035} \\
		& & New & 29 & \multicolumn{1}{r}{41.7} & \multicolumn{1}{r}{21.5} & \multicolumn{1}{r}{.266} & \multicolumn{1}{r}{-.587} & \multicolumn{1}{r|}{.459} \\ \cline{2-9}
		& \multirow{2}{*}{Rev.} & Stand. & 13 & \multicolumn{1}{r}{54.7} & \multicolumn{1}{r}{22.4} & \multicolumn{1}{r}{.027} & \multicolumn{1}{r}{-1.451} & \multicolumn{1}{r|}{.469} \\ 
		& & New & 30 & \multicolumn{1}{r}{58.7} & \multicolumn{1}{r}{20.1} & \multicolumn{1}{r}{-.256} & \multicolumn{1}{r}{-.395} & \multicolumn{1}{r|}{.895} \\ \hline
		
	\end{tabular}
		\vspace{-12pt}

\end{table}

\subsection{Data set preparation}
\label{sec:data_preparation}

In each session, we recorded without pausing the video and audio.
The NASA-TLX questionnaire was answered directly online.
During the data collection process, we took special care not to disturb, or distract, our participants. We manually collected the times when the participant started and ended the visualisation of a given model. Since the answers were given orally, a preparation of that data was also necessary. For the understanding tasks, we had a table with all the elements present in the model, one per column. When listening to the answers, elements that a participant described as being the correct ones were marked with 1, in a row dedicated to each participant. For the reviewing tasks, the procedure was the same, but when the answer was different from the expected, we added a column with that answer, if it was not already present. 
At 
the end, the table contained all the answers given by the participants, and their frequency. Concerning the eye-tracking data, the main areas of the stimulus and its elements were mapped into pixel coordinates to determine which regions and elements the participants were looking at. This allowed tagging the eye-tracking data with the elements being gazed at any given moment, which was a necessary step for computing the eye-tracking metrics used in this paper.

\subsection{Hypotheses testing}
\label{sec:hypotheses_testing}
For testing our hypotheses, we used the Welch \textit{t}-test, instead of the t test, as it is robust to deviations from the normal distribution, different sample sizes and 
variance in the samples, thus following the recommendations on data analysis for Software Engineering empirical evaluations \cite{Kitchenham2016EmSE} (which 
summarises best practices in statistical analysis on other domains).

\begin{figure*}[!htpb]
	\centering
	\begin{subfigure}[b]{0.85\columnwidth}
		\includegraphics[width=\columnwidth]{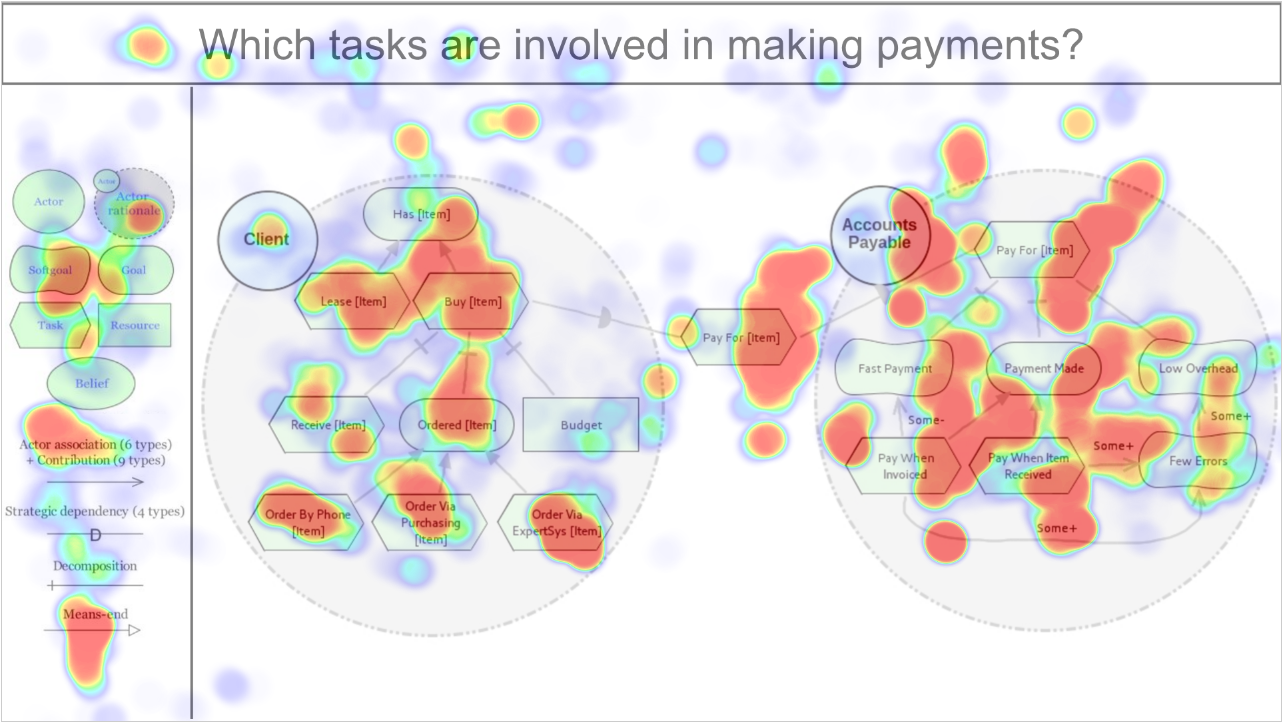}
		\caption{Understand task with standard i* notation}
		\label{fig:originalModel}
	\end{subfigure}
	~
	\begin{subfigure}[b]{0.85\columnwidth} 
		\includegraphics[width=\columnwidth]{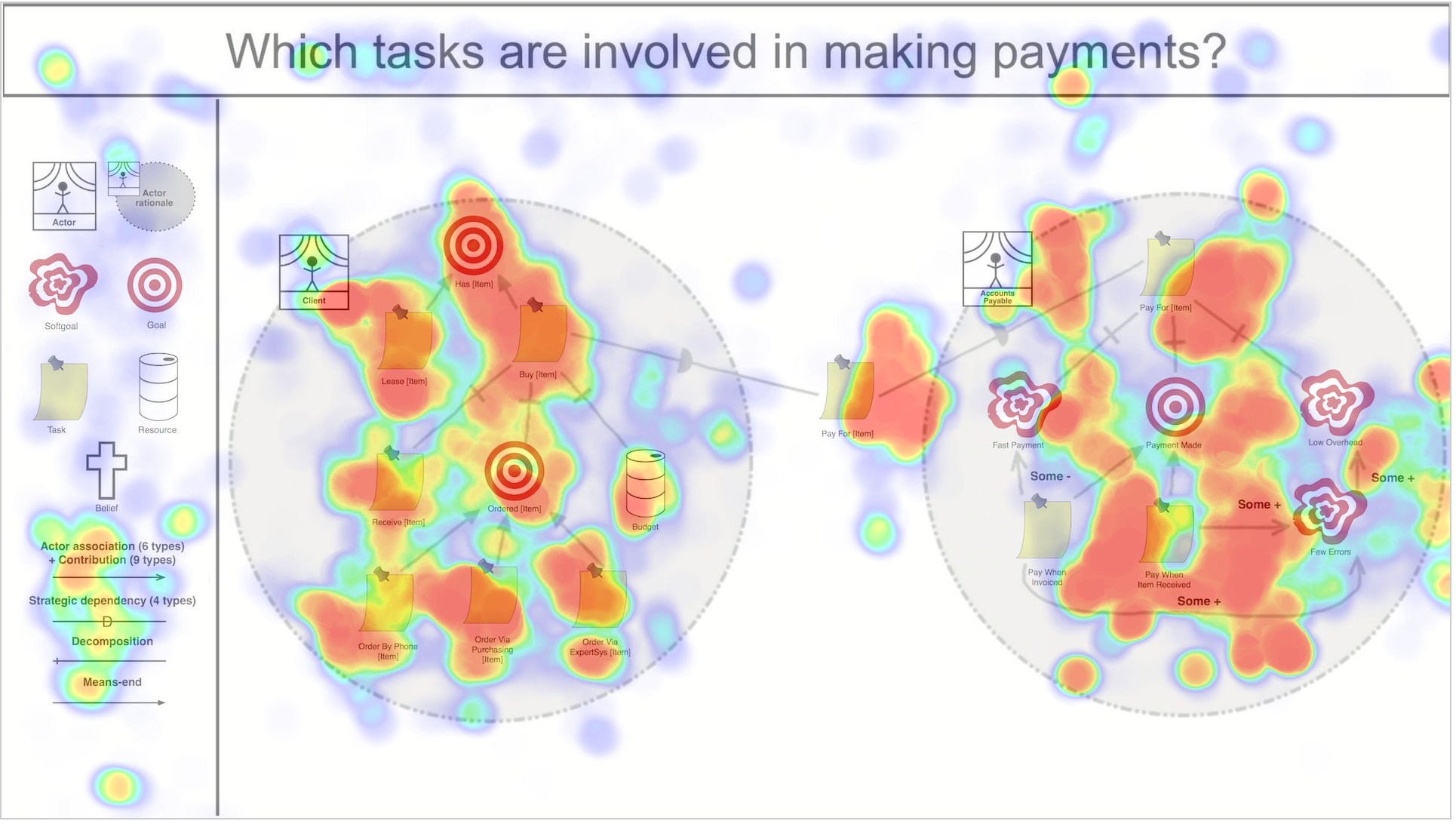}
		\caption{Understand task with the new i* notation}
		\label{fig:newNotationModel}
	\end{subfigure}

\vspace{0.25\baselineskip}	
	\begin{subfigure}[b]{0.85\columnwidth}  
		\includegraphics[width=\columnwidth]{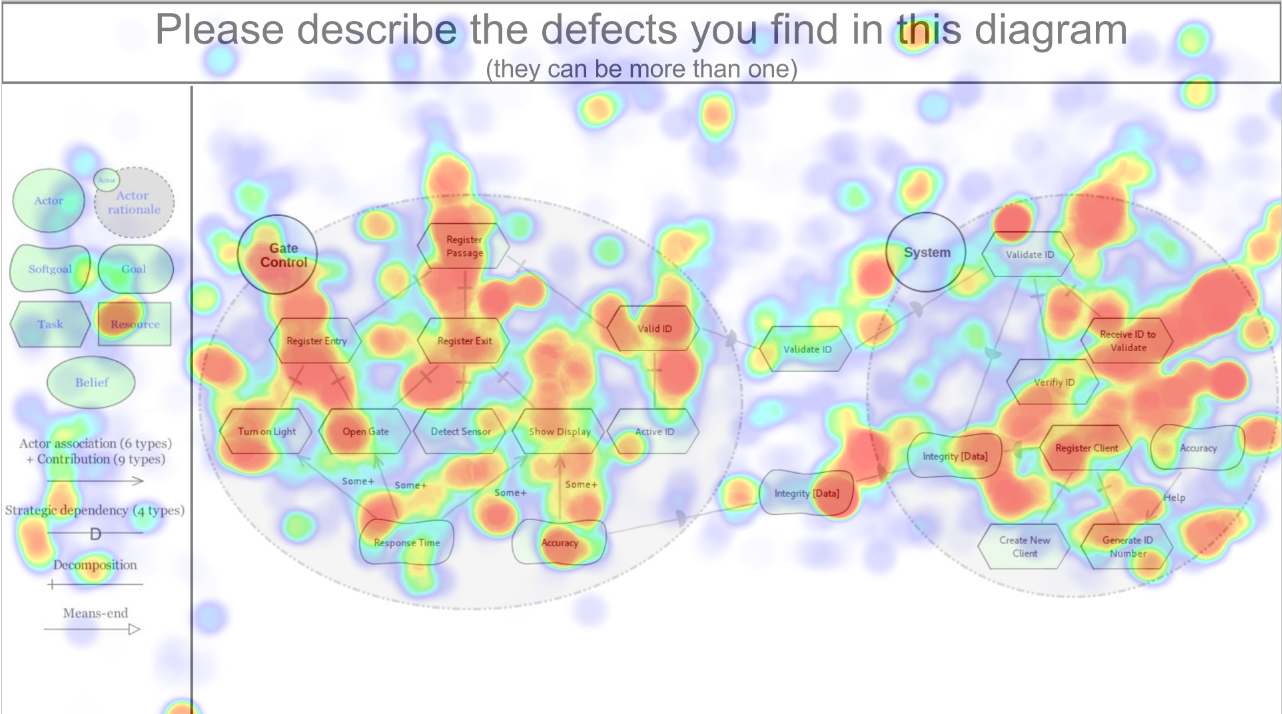}
		\caption{Review task with standard i* notation}
		\label{fig:originalModelDebug}
	\end{subfigure}
	~
	\begin{subfigure}[b]{0.85\columnwidth}
		\includegraphics[width=\columnwidth]{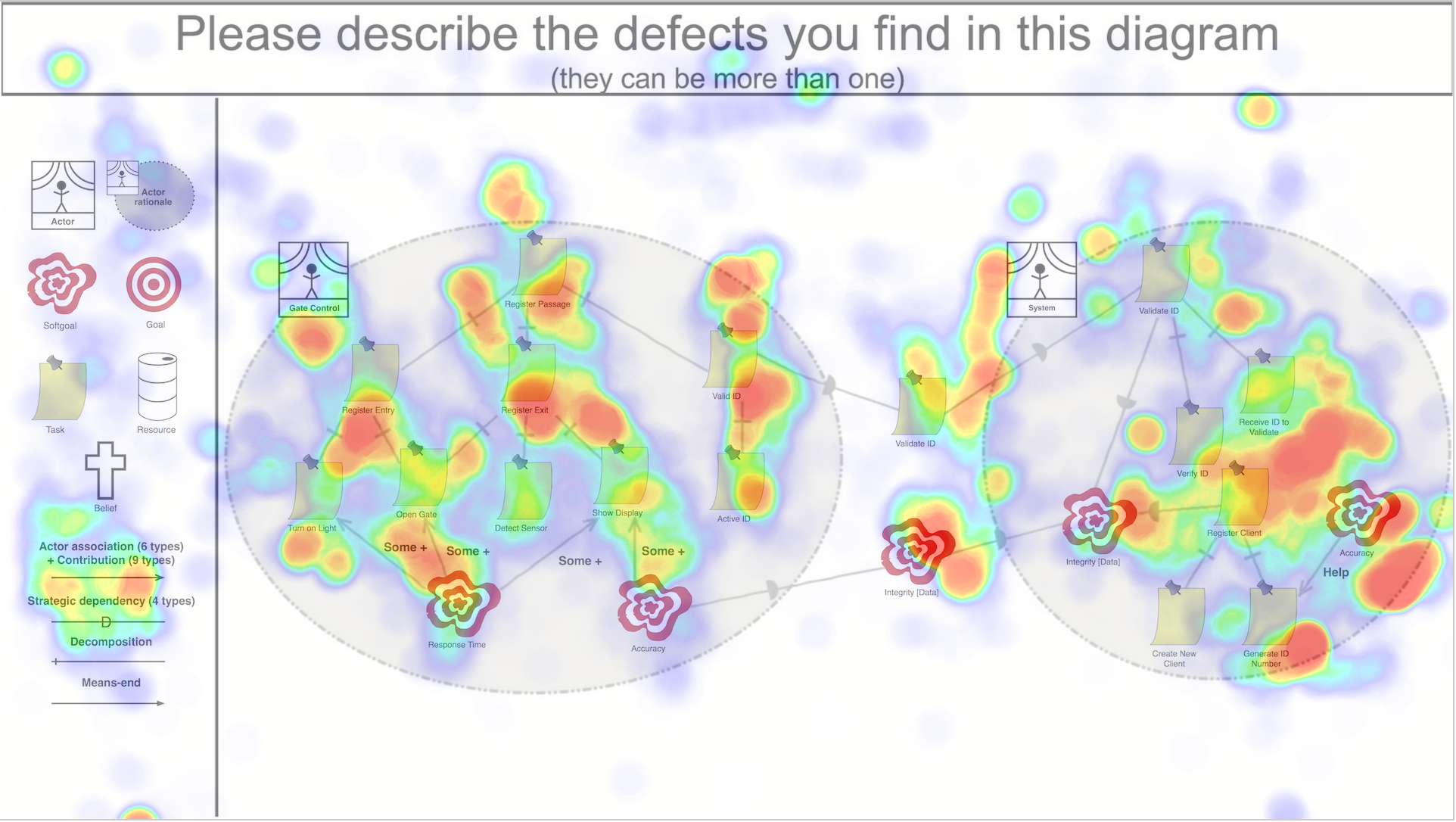}
		\caption{Review task with the proposed i* notation}
		\label{fig:newNotationModelDebug}
	\end{subfigure}
    \setlength{\belowcaptionskip}{-12pt}

	\caption{Heat maps for the understand and review tasks in both notations}
	\label{fig:heatMaps}
\end{figure*}

\textit{RQ1: Does the adoption of a more semantically transparent concrete syntax improve the accuracy, speed and ease when performing understanding tasks on i* SR models?}
Table \ref{tab:WelchTestsUnderstand} summarises the Welch \textit{t}-test results for the \textit{Understand} task. There was a statistically significant difference (see Table~\ref{tab:DescriptiveStats}) between the total number of saccades (\textit{TotSac}) made by participants while understanding the model represented with the standard concrete syntax ($M=47.7$, $SD=12.6$) and those using the new concrete syntax ($M=41.7$, $SD=21.5$; $t(1)=-3.247$, $p=.007$). This suggests a lower visual effort when using the new concrete syntax, in terms of saccades. A similar conclusion can be drawn concerning the number of saccades to the AOI where the language key to the concrete syntax was presented, with a statistically significant difference between the distribution with the standard concrete syntax 
($M=165.4$, $SD=28.7$) and the new concrete syntax 
($M=117.6$, $SD=55.8$; $t(1)=3.469$, $p=.002$). Both variables are related to the \textit{ease} component. 
We found no statistical evidence of differences concerning the remaining variables.
Figs. \ref{fig:originalModel} and \ref{fig:newNotationModel} illustrate the heat maps 
representing the areas more frequently gazed during the understand tasks, with the standard and new concrete syntax, respectively.

\begin{table}[htbp]
	\vspace{-6pt}
	\caption{Welch \textit{t}-test scores for the \textit{Understand} task}
	\label{tab:WelchTestsUnderstand}
\centerline{
	\begin{tabular}{@{}lrrrr@{}}
		\toprule
		\textbf{Metric}    & \textbf{Statistic} & \textbf{df1} & \textbf{df2} & \textbf{Sig.} \\ \midrule
		\textbf{Precision} &  1.186 & 1 & 18.472 & .251 \\
		\textbf{Recall}    &  .316 & 1 & 15.428 & .756 \\
		\textbf{F-Measure} &  .323 & 1 & 15.179 & .751 \\ 
		\midrule
		\textbf{Duration} & -.907 & 1 & 19.254 & .376 \\
		\textbf{FirstDet} & .397 & 1 & 12.669 & .698 \\
		\textbf{LastDet} & .352 & 1 & 12.880 & .730 \\
		\midrule
		\textbf{FixRel} & -.149 & 1 & 11.299 & .884\\
		\textbf{FixIrrel} & .437 & 1 & 11.155 & .670 \\
		\textbf{AvRelDur} & .259 & 1 & 5.302 & .806\\
		\textbf{AvIrrelDur} & .422 & 1 & 18.198 & .678\\
		\textbf{TotSac} & -3.247 & 1 & 12.089 & \textbf{.007} \\
		\textbf{Sac2Key}  & 3.469 & 1 & 30.938 & \textbf{.002} \\
		\textbf{NASA TLX}  & 1.399 & 1 & 42.780 & .243 \\
		\bottomrule
	\end{tabular}}
		
\end{table}

\textit{RQ2: Does the adoption of a more semantically transparent concrete syntax improve the accuracy, speed and ease when performing reviewing tasks on i* SR models?}
Table \ref{tab:WelchTestsReview} summarises the Welch \textit{t}-test results for the \textit{Review} task. Again, there was a statistically significant difference in several variables concerning the \textit{ease} component of cognitive effectiveness, when contrasting the number of relevant fixations, the number of irrelevant fixations, the average duration of irrelevant fixations and the number of saccades to the key. Some of the eye-tracking \textit{ease} metrics suggest a lower complexity (i.e., an easier experience) when using the standard concrete syntax. Others, suggest the opposite. Specifically, the number of relevant fixations using the standard concrete syntax ($M=.86$, $SD=.49$) was higher than the one when using the new concrete syntax ($M=.026$, $SD=.031$; $t(1)=3.935$, $p=.001$). The number of irrelevant fixations has raised from ($M=.282$, $SD=.100$) to ($M=.370$, $SD =.112$; $t(1)=-2.507$, $p=.020$). The average duration of fixations to irrelevant parts of the model was lower with the standard concrete syntax ($M=238.0$, $SD=57.3$) than with the new concrete syntax ($M=292.7$, $SD=103.4$; $t(1)=-2.178$, $p=.036$). Finally, the number of saccades to the language key was lower with the standard concrete syntax ($M=83.8$, $SD=59.1$) than with the new concrete syntax ($M=141.3$, $SD=52.0$; $t(1)=-3.244$, $p=.004$). We found no other statistically significant differences concerning the remaining variables.
Figs. \ref{fig:originalModelDebug} and \ref{fig:newNotationModelDebug} illustrate the heat maps 
representing the areas more frequently gazed during the review tasks, with the standard and new concrete syntax, respectively.

\begin{table}[htbp]
\vspace{-8pt}
	\centering
	\caption{Welch \textit{t}-test scores for the \textit{Review} task}
	\label{tab:WelchTestsReview}
	\begin{tabular}{@{}lrrrr@{}}
		\toprule
		\textbf{Metric}    & \textbf{Statistic}  & \textbf{df1} & \textbf{df2} & \textbf{Sig.} \\ \midrule
		\textbf{Precision} & .000 & 1 & 16.512 & .983 \\
		\textbf{Recall}    & .802 & 1 & 27.868 & .378 \\
		\textbf{F-Measure} & .152 & 1 & 20.143 & .701 \\
		\midrule
		\textbf{Duration}  & -.148 & 1 & 16.967 & .884 \\
		\textbf{FirstDet}& -.133& 1 & 2.985 & .903 \\
		\textbf{LastDet} & .117 & 1 & 5.181 & .911 \\
		\midrule
		\textbf{FixRel} & 3.935 & 1 & 14.769 & \textbf{.001}\\
		\textbf{FixIrrel} & -2.507 & 1 & 22.784 & \textbf{.020}\\
		\textbf{AvRelDur} & .813 & 1 & 28.519 & .423 \\
		\textbf{AvIrrelDur} & -2.178 & 1 & 35.492 & \textbf{.036} \\
		\textbf{TotSac} & .011 & 1& 16.775 & .991 \\
		\textbf{Sac2Key} & -3.244 & 1 & 20.370 & \textbf{.004} \\
		\textbf{NASA TLX}  & .296 & 1 & 20.851 & .592 \\
		\bottomrule
	\end{tabular}
\end{table}

%% file: 7_discussion.tex
\section{Discussion}
\label{sec:discussion}

\subsection{Evaluation of results and implications}
\label{sec:evaluation}
\textit{RQ1: Does the adoption of a more semantically transparent concrete syntax improve the accuracy, speed and ease when performing understanding tasks on i* SR models?}
We found no evidence of improvements brought by the adoption of the new \textit{i*} concrete syntax, in terms of the accuracy and speed with which our participants performed their \textit{understanding} task. The only statistically significant difference observed when contrasting the performance of users with each of the concrete syntaxes conveyed a greater visual effort while using the standard notation, observable through a higher number of saccades in general, and a higher number of saccades targeting the language key, on the left hand side of the screen. Both seem to convey a higher difficulty in using the standard \textit{i*} concrete syntax. That said, the level of success and overall time taken to perform the task are similar, regardless of the particular concrete syntax. Our interpretation is that, even if the particular concrete syntax created some extra difficulties, these must have not been significant. In fact, the self reported perception of the complexity of the task, through the NASA-TLX questionnaire, supports the interpretation that participants evaluated difficulty similarly, in both cases. 

\textit{RQ2: Does the adoption of a more semantically transparent concrete syntax improve the accuracy, speed and ease when performing reviewing tasks on i* SR models?}
As with \textit{RQ1}, we found no evidence of the benefits of the new concrete syntax, when compared to the standard, in terms of speed, or accuracy. Again, there were some 
differences in terms of visual effort. While the effort spent looking at the relevant parts of the model decreased, the effort on looking at irrelevant parts of the model increased, with the new notation. Similar to what we observed for \textit{RQ1}, the feedback provided by the participants through the NASA-TLX questionnaire suggests that, if indeed there was an effort difference, the participants did not notice it. 


\subsection{Threats to validity}
\label{sec:threats}
\textbf{Conclusion validity.} Although we have a reasonable number of participants, higher than most sample sizes reported in other eye tracking experiments (see \cite{sharafi2015systematic}), sample size is a risk, as the results may not apply to larger populations. We plan to extend this study by performing replicas, and we facilitate independent replicas to independent teams, by sharing the materials used in this work.

\textbf{Internal validity.} The potential learning effect for participants from one task 
to the next was mitigated by assigning the tasks to participants in a
way that those 
starting with the understanding task and those starting with the reviewing task were balanced. In addition, participants using one concrete syntax were \textbf{not} using the other one. We found no evidence of learning effects in the data. Finally, special care was taken to guarantee that all the materials produced were easily readable in the 22 inch monitor used for the experiment. We were limited by the technical specifications of the eye-tracker device, such as limitations in the external monitor dimensions and distance to the eye-tracker. The fonts and symbols used had to be big enough for easy visualisation by all participants. As such, the tested models are fragments of larger models. Notwithstanding, presenting only model fragments to focus the attention of the stakeholders is a common technique for improving communication with them. Moreover, our results show that the tasks were already challenging for our participants, with this model size. We need to resolve those technical limitations before the replication with bigger models.



\textbf{External validity.} 
Overall, our participants had little to no prior knowledge in \textit{i*}, making them good surrogates for non-expert stakeholders 
(our target population). Further research is needed to assess how these changes in concrete syntax would impact experienced Requirements Engineers.
Also, the models used in our evaluation are neither representative of all possible alternative concrete syntaxes nor of all \textit{i*} SR models.

\textbf{Construct validity.} Since we have showed a video tutorial
about \textit{i*}, and afterwards participants answered questions about
\textit{i*} models, they might have felt that they were
being evaluated. This 
may have caused an evaluation apprehension
threat, where participants try to look better, which
is confounded to the outcome of the experiment. To mitigate this threat, we have not informed the participants about what was being tested, i.e., their accuracy, speed and ease in the performed tasks.

\subsection{Inferences}
\label{sec:inferences}
Inferences are discussed contrasting the results of the standard concrete syntax with the new \textit{i*} concrete syntax.

\textbf{Similar speed and accuracy.} Our results suggest that 
for \textit{i*} models of the complexity 
used in this evaluation 
there was no observable benefit in the \textit{speed} and \textit{accuracy} with which the participants were able to conduct their tasks. 
Two possible explanations for cancelling the effect of using symbols with a greater semantic transparency are: 
\begin{inparaenum}[(1\upshape)] \item the presence of a language key that facilitates the interpretation of the symbols in such a way that the higher semantic transparency of the new concrete syntax has no effect in the results, and \item the results were mostly influenced with the difficulties of our participants with semantic aspects of the models rather than with syntactic ones. \end{inparaenum}
Concerning \textit{ease}, we did find some indicators of eye-tracking suggesting different visual efforts. Further research is necessary to assess how consistently these results occur with other users, models and concrete syntaxes.

\textbf{No deep overall impact of visual effort.} 
The visual effort is lower for the new concrete syntax, as participants seem a bit more ``lost'' with the standard concrete syntax, making a more erratic model navigation (see the more scattered heat map footprint in Fig. \ref{fig:heatMaps}). 
However, this was not perceived as a shortcoming by the participants. They were not even aware that they struggled more with the navigation, as suggested by their answers to the NASA-TLX questionnaire. Thus, although navigating in the standard concrete syntax models was visually harder, this had no practical impact in their overall performance. If the tasks were longer or in a higher number, though, the results could have been different, due to fatigue. This should be explored in subsequent studies.


\textbf{Better symbol semantic transparency did not imply better model understanding.} 
This is somewhat in line with the findings in \cite{vanDerLinden2018TSE}, reporting that the application of the PoN theory is complex, often leading to sub-optimal concrete syntaxes proposals and evaluations. Even when the semantic transparency of the concrete syntax significantly improves, this does not necessarily translate into better performance when using the models, due to the context provided by the model, and, when available, the presence of a language key. Furthermore, semantic transparency is just one of the 9 PoN principles. Hence, we suggest that future studies consider syntactic improvements based on more than a single PoN principle. Plus, more realistic scenarios should be considered. 


%% file: 8_related_work.tex
\section{Related work}
\label{sec:related_work}

Several studies were performed upon different 
modelling languages, particularly UML, BPMN, and some 
goal-oriented languages, such as \textit{i*} and KAOS. These studies aim at detecting problems concerning the languages' concrete syntax by using the PoN 
set of principles, and propose solutions to mitigate them. 
Moody et al. \cite{moody2008evaluating} propose several improvement recommendations for the concrete syntax of diagrams defined in UML 2.0 
while Kouhen et al. \cite{el2015semantic} evaluate 
UML with a set of experiments and report on its lack of semantic transparency. 
Genon et al. \cite{Genon:2010:ACE:1964571.1964605} evaluate the cognitive effectiveness of the BPMN 2.0 concrete syntax, and Moody \cite{moody2011diagram} identifies in BPMN serious issues that may hinder its usability and effectiveness in practice, particularly for communicating with end users. Regarding goal-oriented approaches, Moody et al. \cite{Moody2010REJ} analyse 
the cognitive effectiveness of \textit{i*}, Caire et al. \cite{Caire2013RE} propose an approach to designing concrete syntaxes, demonstrated with \textit{i*}, that actively involves novice users in the process. Matulevičius et al. \cite{matulevivcius2007visually} evaluate how KAOS and Objectiver, its tool, help the modelling activity, 
offering recommendations for modellers, language designers and tool developers.

Störrle \cite{Storrle2011VLHCC} studies the impact of the usage of good \textit{vs.} bad diagram layouts on model comprehension tasks when using UML, in particular use cases, class, and activity diagrams. On a similar research line, Santos et al. \cite{Santos2016RE} evaluate the effect of the layout guidelines on the \textit{i*} models understandability, by using eye-tracking. Other studies with eye-tracking, assessed the effort involved in the comprehension of software models like BPMN \cite{Petrusel2013CAiSE}, ER \cite{cagiltay2013performing}, or TROPOS \cite{Sharafi2013ICPC}.

Albeit their importance, several 
PoN studies focused on the evaluation of individual 
symbols, and on the stakeholders' ability to correctly recognise them. 
Yet, software engineers use models. 
A significant difference from previous studies to this paper is that we perform our evaluation at the model level, rather than through isolated symbol recognition tasks.


%% file: 9_conclusions.tex
\section{Conclusion}
\label{sec:conclusion}

We performed a quasi-experiment to compare the \textit{accuracy}, \textit{speed}, and \textit{ease} of the standard \textit{i*} concrete syntax and 
an alternative \textit{i*} concrete syntax that resulted from the most successful symbol recognition evaluations for \textit{i*} \cite{Caire2013RE}. 
A total of 57 participants 
performed understanding and reviewing tasks on \textit{i*} SR models.
The data collected showed that the 
alternative concrete syntax had no significant impact in the \textit{accuracy} and \textit{speed} with which 
participants conducted their tasks. Increased semantic transparency alone did not lead to a better performance with the new \textit{i*} concrete syntax. 
The presence of a language key and the context provided by the model may have mitigated the effect of the increased semantic transparency of the new \textit{i*} symbols. 
For \textit{ease}, we found some indicators of eye-tracking suggesting different visual efforts.

We only addressed one of the nine PoN principles in this study.
Further studies should consider the various principles, the interactions among these, as well as their influence on the actual performance of practitioners in understanding and reviewing social goal models.
%
It would be interesting to understand if the new concrete syntax has any drawback (e.g., in model construction) that hinders 
performance, 
or why the NASA-TLX questionnaire results do not support the visual effort clear in the heat map, or still, understand the fixation time on relevant/irrelevant AOIs and how they differ between the two groups of participants.
Finally, it is necessary to assess how consistently our results occur with other users, models and concrete syntaxes. 
We plan to replicate the experiment in other contexts, and apply it to bigger and more complex models.
